\documentclass[letterpaper, 10 pt, conference]{ieeeconf}  
\IEEEoverridecommandlockouts                         
\usepackage{cite}
\usepackage{graphics} 
\usepackage{epsfig} 
\usepackage{hyperref}

\title{\LARGE \bf
Massive RF Simulation Applied to School Connectivity in Malawi
}

\author{Ermanno Pietrosemoli, Marco Rainone, Marco Zennaro and Chomora Mikeka} 

\begin{document}

\maketitle
\thispagestyle{empty}
\pagestyle{empty}

\begin{abstract}
Providing Internet connectivity to schools has been identified as paramount for development, for instance in the Giga project, cosponsored by ITU and UNICEF, with the goal of connecting every school to the Internet by 2030.

For a country wide deployment, it is imperative to perform a thorough  planning of the whole installation, using radio frequency (RF)  propagation models. While statistical models based on empirical RF propagation data gathered in different scenarios can be employed, for point to point links at microwave frequencies the existence of a clear line of sight (LOS)  is  normally a prerequisite. 

The Irregular terrain model which makes use of digital elevation maps (DEM) has proved quite effective for simulating point to point links, but its  application to a great number of links becomes  time consuming, so we have developed  an automated framework to perform this task. As a case study we have applied this framework  in the planning of a project mired at providing connectivity to primary and secondary schools all over the country of Malawi.

\end{abstract}

\section{INTRODUCTION}
Affordability of Internet connectivity for rural schools in developing countries is still an open issue, with profound impact in the quality of education.
Fiber optic is not cost effective for last mile in sparsely populated areas, so wireless is the preferred technology. Satellites are well suited for places far away from terrestrial infrastructure, but terrestrial solutions are normally less expensive \cite{alt-n}. TV White space \cite{Mla}, \cite{TVWS},  \cite{TVWS-M}, has not met the expectations of price reduction but microwave links have been widely deployed by  wireless Internet service providers (WISP) in many countries.

Cellular operators find it difficult to obtain a reasonable return of investment in  sparsely populated area, so modified WiFi for long distance using inexpensive commercial hardware \cite{flickenger2008very} has been a choice technology for serving remote areas.

Efforts to leverage existing cellular towers  to offer subsidised services for educational institutions have been attempted in many countries, normally relying on wireless links planned using the Hata-Okumura RF  propagation model \cite{hata-oku} to asses coverage. This model has proved adequate in urban environments for short  distance links. In scarcely populated areas the distances are much greater, and the influence of the terrain profile is the dominant factor in determining coverage. 

Authors in \cite{model-compar} performed a comparison between the results of some of the most commonly used propagation model against field measurements and concluded that the Longley-Rice (LR) \cite{LR} model errors were lower than those of the Hata and ITU-R P.1536 models, but involved more planning complexity. Our paper  focuses specifically in reducing the complexity in applying the LR method.
Splat! \cite{splat} is an open source implementation of the ITM model that we have used to build the BotRf \cite{bot} tool that facilitates the analysis of point to point (PtP) links. 
Although splat! and similar programs like the widely used radiomobile \cite{rmob} can simulate the coverage of a point to multipoint (PtMP) station in terms of different colors, this is not good enough to ascertain the feasibility of serving a specific spot from a given tower.

Therefore, we have developed a framework to automate the procedure of determining the terrain profile between a given tower and a large number of schools within a specific radius, and from that profile verifying if there is  an unencumbered line of sight (LOS). In the 5 GHz unlicensed frequencies that are the most cost effective to provide connectivity at long distances, a clear line of sight means that a link is feasible. Even a single obstacle might not preclude connectivity in some cases. 

It is also worth noting that sometimes a  further apart tower from a given school might be used to circumvent a terrain obstacle laying on the path to the nearest tower to achieve connectivity. The system we have developed assesses the existence of LOS from each school to nearby towers using as inputs a list with the coordinates of the schools and another list with the coordinates of the usable towers.  

As a case study, we have applied this methodology in the country of Malawi in which we have previously deployed communication networks \cite{mw}, \cite{mw8}.

\section{Architecture of the proposed network solution}
\begin{figure}[htbp]
	\centerline{\includegraphics[width=90mm,scale=1.5]{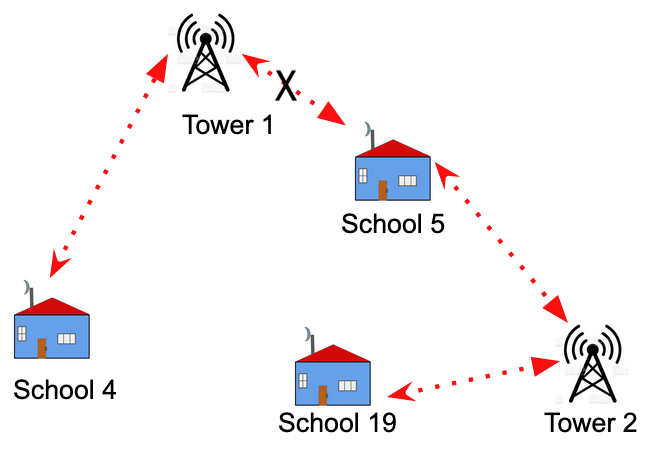}}
	\caption{Schematic layout. School 5 is closer to Tower 1, but because of a terrain obstacle it is being served by Tower 2}
	\label{layout.png}
\end{figure}

The system layout,  shown in Fig. \ref{layout.png}, is based on establishing a Point of Presence (POP) in each of the telecommunications towers which are already connected by fiber optics and where stable electrical power is normally provided. Where no fiber is available, backhaul can be provided by a terrestrial microwave  or  satellite link. Schools at distances of up to 50 km can be connected to the towers via 5 GHz modified WiFi links.

\subsection{Tower Infrastructure}

In each  tower  four point to multipoint radios,  fitted with an integrated sectorial antenna with a 90º coverage area, will be installed. Given the low cost of radios with integrated antennas, this arrangement provides higher antenna gain and higher throughput as compared with using a single radio with an omnidirectional antenna, without incurring in a severe cost penalty. 

\subsection{School Infrastructure}

 For every  the schools to be served, a single point to point  5 GHz radio for connecting to the donor tower will be enough. For short distances an omnidirectional antenna will suffice while for longer distances a high gain antenna will be required.
The radio will be connected to a 2.4 GHz WiFi Access Point which will provide Internet to the school's students and staff. 
In most cases a simple pole or existing support structure can be used to attach the radio with its antenna. Where necessary, a small  solar photovoltaic system can be installed to supply the required electrical power.

\begin{figure}[htbp]
	\centerline{\includegraphics[width=90mm, scale=1.5]{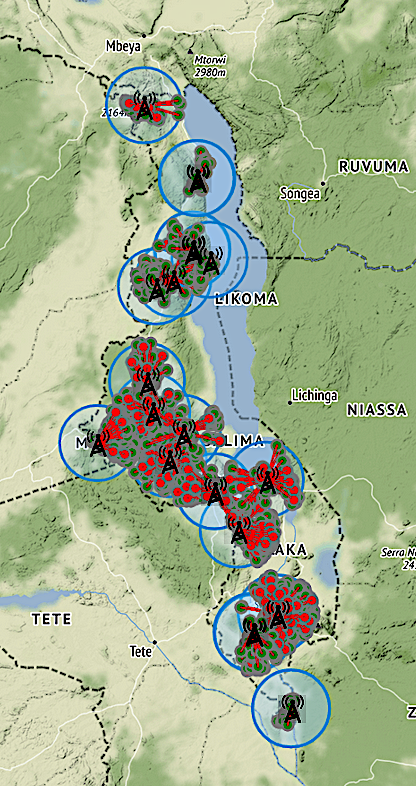}}
	\caption{Telecommunication Towers in Malawi showing 50 km radius circles  of possible coverage. Red lines represent links to schools.}
	\label{MalawiT.png}
\end{figure}

\begin{figure}[htbp]
	\centerline{\includegraphics[width=90mm, scale=1.5]{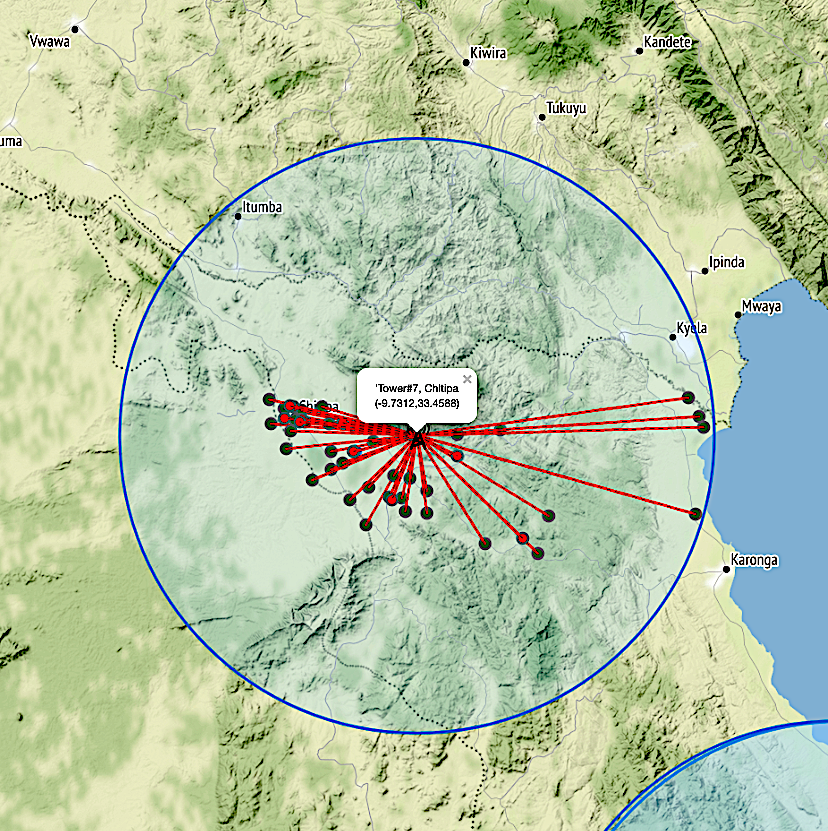}}
	\caption{`Coverage of Tower 7, northernmost  in Malawi, showing several secondary (red dots)  and primary (green dots) schools  contained in a 50 km radius circle.}
	\label{Tower7.png}
\end{figure}

\section{Platform developed}

We have developed a series of Python programs, based on the open source splat! tool, to simulate PtP links on a massive scale and to visualize the results in a friendly way using a web application. The user has to provide as input two lists with the coordinates of the towers and those of the schools to be connected. In the case of towers the height above ground should also be specified. As schools are typically one or two stories high, a height of 10 meters for the antenna is provided by default.

\begin{figure}[htbp]
	\centerline{\includegraphics[scale=0.3]{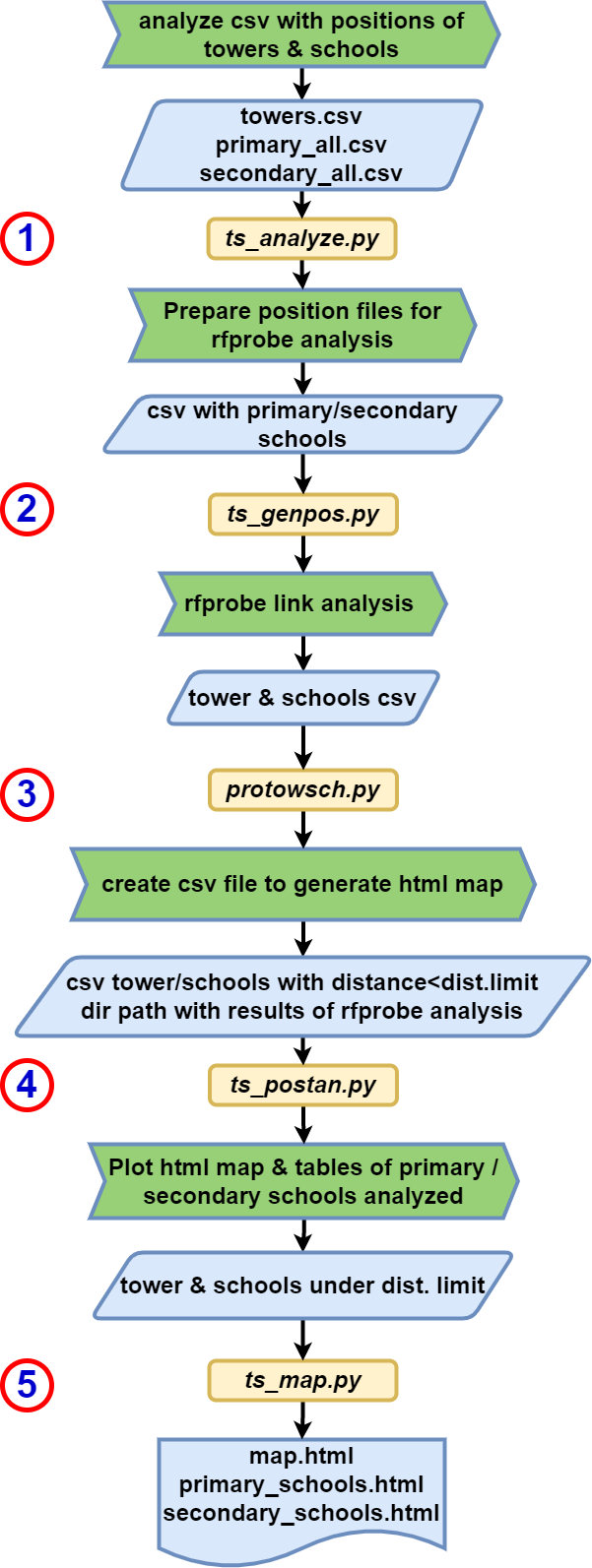}}
	\caption{Workflow of Python tools developed}
	\label{tosc01.png}
\end{figure}

The logical workflow of the tools is shown in Fig. \ref{tosc01.png}:

 \begin{enumerate}
\itemsep=0pt
\item  Analyze the coordinates of the towers and of the schools to determine the tower to be used to service a given school. This is performed by means of the ts\_analyze.py Python script that prepares the files for further analysis.   
\item  Apply ts\_genpos.py to process the  data in order to generate the positions of both ends the link
\item  Leverage the ITM model and digital elevation maps by means of the BotRf \cite{bot} tool based on the open source splat!  \cite{splat} program with  protowsch.py to generate the terrain profiles as shown in Fig. \ref{Profile.png}
\item  Run ts\_postan.py to plot html maps and tables of schools  
\item  Apply ts\_map.py to generate the overall map.html shown in Fig. \ref{MalawiT.png}
\end{enumerate}

 \section{Malawi case study}

Malawi is an a African country in which many of the schools in rural areas are not yet connected, despite efforts to apply solutions like those based in TV White Space \cite{TVWS},\cite{TVWS-M}. The Malawian  government was able to obtain the use of a number of telecommunication towers, which already have fiber optics connectivity, to provide Internet access to primary and secondary schools across the country. There are over 2k schools to be connected. A map showing the locations of the schools and that of the towers is shown in Fig. \ref{MalawiT.png}. The blue circle indicates the 50 km radius around each tower inside which  schools might potentially be served with a meaningful connection.

Armed with the coordinates of these towers and those of the schools, we applied our automated procedure to determine the feasibility  of serving surrounding schools from each of the towers. 

The first step is to find those links within a 50 km radius from the tower that have a  clear  line of sight, which in most cases guarantees that the microwave link will work. Then we also examined the cases in which there is a single obstacle, which sometimes can be overcome if the link is short. If none of these conditions is met, we try establishing  links towards other towers.

\begin{figure}[htbp]
	\centerline{\includegraphics[scale=0.4]{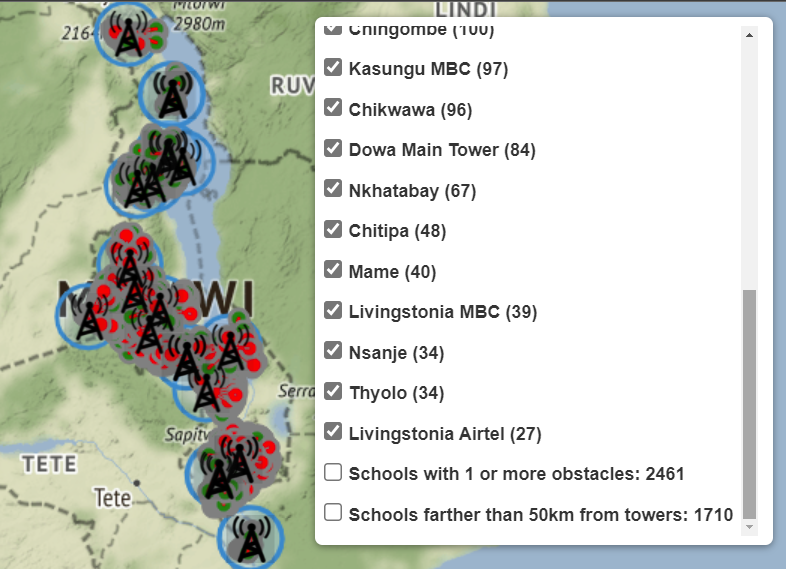}}
	\caption{List of towers and the number of schools potentially covered}
	\label{menu-destra.png}
\end{figure}

Applying the tool to the list of towers and schools in Malawi will produce Fig. \ref{MalawiT.png}. Clicking on the stack symbol in the upper right corner will show a list of the telecommunication towers and the number of schools that have a clear LOS  to them. It  can also show the schools with one or more obstacles in their path to the tower as seen in Fig. \ref{menu-destra.png}.
Clicking over any of the towers will  reveal its coordinates  along with red lines to the schools  within a 50 km radius, as shown in Fig. \ref{Tower7.png}. Green dots correspond to primary schools and red dots to secondary ones.

\begin{figure}[htbp]
	\centerline{\includegraphics[width=90mm, scale=1.5]{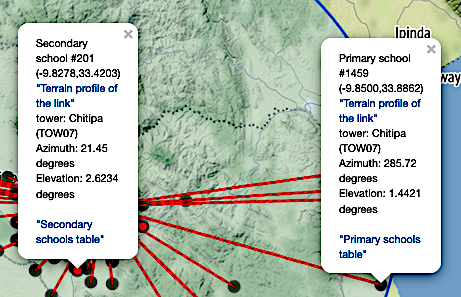}}
	\caption{Clicking over any school shows its identification, latitude, longitude, elevation, and the azimuth and elevation angles towards the tower. It also reveals   links  to the terrain profile between the school and the tower and to a table containing detailed information of all schools in the category, specifying which ones are within 50 km and if they have  a clear LOS to the tower.}
	\label{PopUp.png}
\end{figure}

\begin{figure}[htbp]
	\centerline{\includegraphics[width=90mm, scale=1.5]{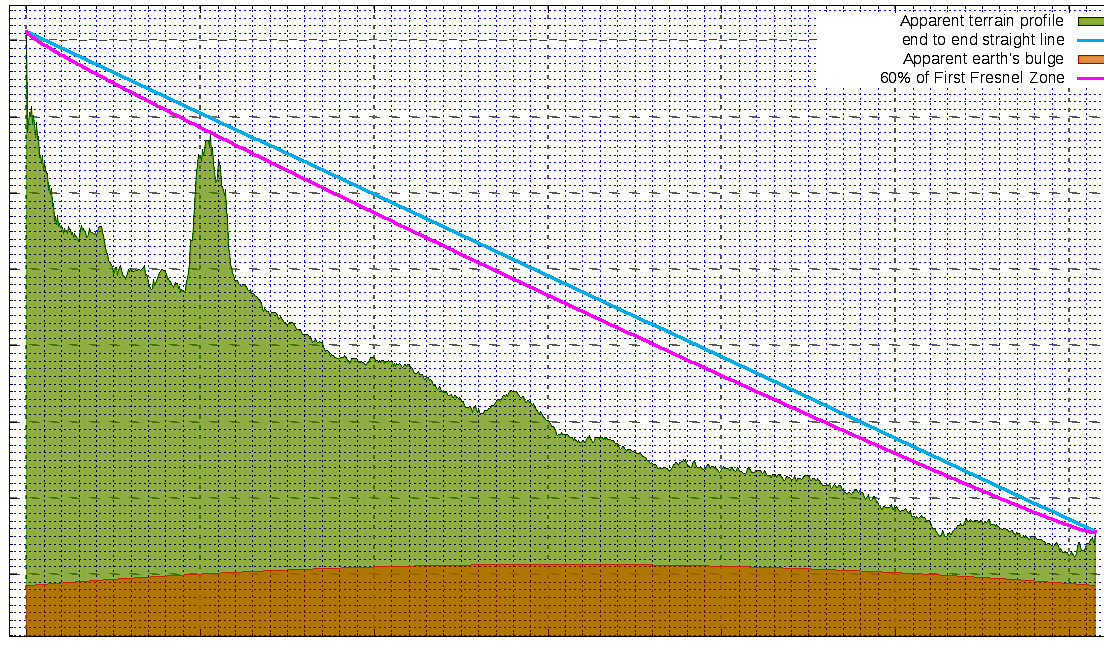}}
	\caption{Terrain profile between primary  school 3140 and  tower 7, showing height above ground  (green) versus distance (end to end 31 km). The upper blue line is the line of sight between the end points, the red curve represents 60 \% of the Fresnel zone at 5 GHz and the brown zone depicts the  curvature of the earth.}
	\label{Profile.png}
\end{figure}

Clicking on any of the dots opens a pop up window,  shown in Fig. \ref{PopUp.png}, that reveals the id and coordinates of the school as well as the azimuth and elevation towards the tower, data used for aiming directional antennas. There are also links  to "Terrain profile of the link" Figure \ref{Profile.png}, and to "Primary or Secondary  school tables" which list their id, coordinates, azimuth and elevation towards the tower and whether there is LOS or not.

Running the tool to simulate the links for 2k schools and to produce the maps and profiles required 24 hours of CPU time on a powerful server.

\section{CONCLUSIONS AND FUTURE WORK}
\label{sect5}
We have presented a tool for the planning of  affordable Internet access for schools in remote areas, employing low cost commercially available RF equipment  that use unlicensed frequencies. A case study addressing the connection of schools in Malawi has been proposed and shows how open source tools can be used to plan wireless links on a massive scale. WiFi has proven to be a reliable wireless technology, currently connecting more than 20 billion devices worldwide.  We believe that modified WiFi for long distances  can offer an affordable solution to provide meaningful connectivity to schools by making use of existing telecom towers. Concerns about interference with already deployed radios are mitigated by the  difference in frequencies bands exploited.
We look forward to undertaking a similar simulation for other countries in the future.


\end{document}